\documentclass[aps,a4paper,superscriptaddress,preprintnumbers,showpacs,amsmath,amssymb]{revtex4}
\usepackage{graphicx}
\usepackage{dcolumn}
\usepackage{color}
\usepackage{latexsym,amsfonts}
\usepackage{textcomp}
\usepackage{bm}

\usepackage{ulem}

\def\beq{\begin{equation}}
\def\eeq{\end{equation}}

\begin{document}

\title{Unification of gravity and quantum field theory from extended noncommutative geometry\footnote{This paper bas been published in Mod.Phy.Lett.A 32 (2017) 1750030.}}

\newcommand*{\PKU}{School of Physics and State Key Laboratory of Nuclear Physics and Technology, Peking University, Beijing 100871,China}\affiliation{\PKU}
\newcommand*{\CICQM}{Collaborative Innovation Center of Quantum Matter, Beijing, China}\affiliation{\CICQM}
\newcommand*{\CHEP}{Center for High Energy Physics, Peking University, Beijing 100871, China}\affiliation{\CHEP}

\author{Hefu Yu}\email{hefuyouxiang@pku.edu.cn}\affiliation{\PKU}
\author{Bo-Qiang Ma}\email{mabq@pku.edu.cn}\affiliation{\PKU}\affiliation{\CICQM}\affiliation{\CHEP}

\date{\today}

\begin{abstract}
We make biframe and quaternion extensions on the noncommutative geometry, and construct the biframe spacetime for the unification of gravity and quantum field theory. The extended geometry distinguishes between the ordinary spacetime based on the frame bundle and an extra non-coordinate spacetime based on the biframe bundle constructed by our extensions. The ordinary spacetime frame is globally flat and plays the role as the spacetime frame in which the fields of the standard model are defined. The non-coordinate frame is locally flat and is the gravity spacetime frame. The field defined in both frames of such ``flat'' biframe spacetime can be quantized and plays the role as the gravity field which couples with all the fields to connect the gravity effect with the standard model. Thus we provide a geometric paradigm in which gravity and the quantum field theory can be unified.
\end{abstract}
%02.40.Gh Noncommutative geometry
%04.60.-m Quantum gravity
%11.10.Nx Noncommutative field theory
%12.10.-g Unified field theories and models

\maketitle

\section{Motivation}
\label{sec0}
The unification of gravity and quantum field theory (QFT) is an important goal of physics research. Gravity is described by Einstein's general relativity with curved spacetime, whereas the quantum fields in the standard model
are quantized in the flat spacetime. Both theories have achieved overwhelming successes to almost all physical observations related to gravity, electromagnetic, weak and strong interactions, though it is still obscure as to how to unify the two theories in one framework. The unification contains essentially two ingredients: one is the quantization of gravity, and the another is to include the gravity effect in the gauge field theory of the standard model. Many attempts have been devoted to quantize both gravity field and standard model fields in curved spacetime. Recently there is a novel proposal to treat the gravity force
in the QFT framework with biframe spacetime~\cite{wu.qg}. The biframe space is flat in the sense that the coordinate spacetime frame is globally flat and the non-coordinate spacetime frame is locally flat. Quantum fields of the standard model is defined in the coordinate spacetime frame, while the gravity field (named as gravifield in Ref.~\cite{wu.qg}) is sided in both of the two frames and couples with all the fields to make the standard model meet general relativity. In such flat biframe spacetime the gravifield can be quantized as the quantum fields in QFT. Thus the biframe spacetime theory provides a viable way to quantize gravity in a similar manner as the quantization of fields in the standard model.

The purpose of this work is to establish the biframe spacetime in the algebraic framework of the noncommutative geometry~\cite{con,nc.rev}. We make extensions on the geometry such that the frame bundle, which corresponds to ordinary spacetime frame, of the manifold can be extended to a ``biframe bundle'', which corresponds to an extra non-coordinate spacetime frame. The geometric features (including symmetries and in which sense the space is flat) in the two frames of the biframe bundle are distinguished. Thus the bundle can be flat in the sense that the coordinate frame is globally flat while the non-coordinate frame is locally flat at every point of the bundle. The standard model is expressed in the coordinate frame corresponding to the ordinary frame bundle of the manifold. The general relativity is fulfilled (effectively at low energy) in the biframe spacetime. The field defined in both of the two frames plays the role as the gravifield and connects the standard model with general relativity. This gravifield can be quantized in the flat biframe spacetime. Thus we achieve at an unification of gravity and QFT in a geometric paradigm.

Noncommutative geometry theory provides a promising framework to unify the standard model and general relativity~\cite{nc.rev,con,Lizzi:1996vr,Chamseddine:1991qh,Connes:1996gi,Chamseddine:2006ep,Chamseddine:2010ud}. Physical notions of the geometry are encoded by the spectral triple $(\mathcal{A}, \mathcal{H}, D)$, where $\mathcal{A}$ is the involutive unital algebra represented on the Hilbert space $\mathcal{H}$ and $D$ is the Dirac operator in $\mathcal{H}$~\cite{Chamseddine:1991qh,Connes:1996gi,Chamseddine:2010ud,nc.rev}. In the geometric framework, the Einstein action and the bosonic part of the action of the standard model are expressed by the trace purely relative to even order products of $D_A$ (the Dirac operator corrected by inner fluctuations, which is introduced later), and the fermionic part of the action of the standard model is reinterpreted by $\frac{1}{2}\langle J\tilde{\xi}|D_A|\tilde{\xi}\rangle$ (where $\tilde{\xi}$ is the Grassmannian analogue of $\xi\in\mathcal{H}$). In this sense, general relativity and the standard model are expressed in one framework. The space is quantized by quantizing the chirality operator, which is considered as the higher degree Heisenberg commutation relationship involving algebras %(of real or complex functions of coordinates)
that play roles of coordinate spaces~\cite{Chamseddine:2014nxa,Chamseddine:2014uma}. How to characterize the quantum gauge fields affected by the quantized space, and is there an unified quantization process for gravity and the fields in the standard model, are still open questions in the geometric framework.

The recent work in Ref.~\cite{Chamseddine:2016pkx} unifies gravity and QFT by describing the geometry in SO$(1,13)$ symmetry group and restricting the gauge part to SO$(10)$ grand unified theory. We provide another viewpoint for the above questions by extending the geometry in the minimal (i.e., almost non-effective) way. Namely, we make two extensions on the geometry, and each of the extensions alone keeps the physical action invariant. Only the combination of the two extensions has physical effects. The first extension is the biframe extension on the spectral triple of the geometry. Under this extension, the spectral triple is divided into the part encoding the ordinary frame bundle and the part encoding the non-coordinate part of the biframe bundle. Physical notions encoded by the spectral triple are then defined in two frames, i.e., the coordinate frame and the non-coordinate frame. In usual conditions, the non-coordinate frame and the fields in such frame do not give explicit contribution to the action. The second extension is the quaternion extension on the fields. This extension combined with the biframe extension makes the fields in the two frames be governed by different and proper symmetries. Then field defined in both the coordinate frame and the non-coordinate frame is governed by the global Lorentz symmetry and the local spinnic Lorentz symmetry SP$(1,3)$. This field is equivalent to the gravifield $\chi$ sided on the biframe spacetime in Ref.~\cite{wu.qg}. The spectral action of such geometry characterizes a unified theory for gravity and QFT.

\section{Biframe extension on the geometry $M$}
\label{sec1}
The fundamental nature of the noncommutative geometry is determined by an even real spectral triple, i.e., $(\mathcal{A}, \mathcal{H}, D)$ endowed with a $\mathbb{Z}/2$-grading operator $\gamma$ ($\gamma^2=1$) and a real structure of antilinear isometry $J$~\cite{Chamseddine:1991qh,Connes:1996gi,Chamseddine:2010ud,nc.rev}. The geometry is based on the product space $M\times F$, where $M$ is the Riemannian manifold whose spectral triple is $(C^\infty(M),$ $ L^2(M,S),$ $\partial\!\!\!/_M,$ $\gamma_M,$ $J_M)$ and $F$ is a finite geometry~\cite{nc.rev,con}.

One can discuss the geometry of $M$ by not only a tangent bundle, but also a $G(x)$-principal bundle $P(M)$ with $G(x)$ defined as the general linear group (which is locally compact) of $\varsigma(x)\in C^\infty(M)$. This principal bundle associated with proper vector field space $E(x)$ is a frame bundle:
\begin{equation}\label{fra.bun}
  F(M)=P(M)\times E(x)= \{(G(x), e_\mu)\}=\{(\varsigma(x), \partial_\mu)\},
\end{equation}
where $\mu \in \{0,1,2,3\}$. We mention that fibers $G(x)$ of $P(M)$ are homeomorphic to a ``standard fiber'' $G$:
\begin{equation}\label{G}
\varphi_x: G(x) \rightarrow G,
\end{equation}
while basis of the representations of $G$ can be defined as (a permutation of) linearly independent functions of $G(x)$.%, and $G$ can be represented as operations on (or permutations of) the basis.
We denote the element of $G$ as $\varsigma(1)\equiv \varphi_x \varsigma(x)$, any of which is homeomorphic to some coordinate dependent $\varsigma(x) \in C^\infty(M)$.

In this work, we consider a special case that the group of the principal bundle $BP(M)$ is $G(x)\times G$, while the vector space associated to the corresponding frame bundle $BF(M)$ is $E(x)\times E$. Here $E$ is a non-coordinate frame corresponding to fields of the non-coordinate group $G$. Without extra assumptions, one gets the minimal case, i.e., $E=\{e=1\}$. Then fields $g\in G$ associated with such frame can be written as $g^a e_a=g$ ($a\in \{1\}$). Since $BF(M)$ corresponds to not only the ordinary coordinate spacetime frame but also an extra non-coordinate spacetime frame, we call $BF(M)$ a ``biframe bundle''. In usual conditions, as mentioned, the non-coordinate frame is hidden, i.e., has no physical effects, and fields associated with this frame can ignore the label $a$ of this frame, i.e., $g^a\equiv g \in G$.

We introduce the biframe bundle $BF(M)$ by biframe extension on ingredients of the spectral triple of $M$. We first extend the Dirac operator to
\begin{equation}\label{ex.par}
  \triangle\partial\!\!\!/_M=\frac{1}{2}(\partial\!\!\!/_M\otimes 1 + 1 \otimes \partial\!\!\!/_M).
\end{equation}
This extended Dirac operator distinguishes the parts of $BF(M)$ defined in the coordinate frame between those defined in the non-coordinate frame. This is because the Dirac operator, modulo a gauge transformation, is the inverse of the Euclidean fermion propagator~\cite{con}, and the local nature of the physical framework is encoded by the (nontrivial part) of the Dirac operator and fields coupling with this operator. Thus we can consider the fields coupling with the $\partial\!\!\!/_M$ in $\triangle \partial\!\!\!/_M$ as those restricted to coordinate frame of $BF(M)$, and consider the fields coupling with the ``1'' in $\triangle \partial\!\!\!/_M$ as those restricted to the non-coordinate frame of $BF(M)$.

Then we extend $\varsigma(x)$ in the spectral triple to
\begin{equation}\label{bi.f}
  \triangle \varsigma(x)=\varsigma \otimes \varsigma,
\end{equation}
where $\varsigma$ is defined as
\begin{equation}\label{f.x.1}
  \varsigma=\left\{\begin{array}{cc}
  \varsigma(x), & \mbox{when coupling with $\partial_\mu$ in $\triangle \partial_\mu$,} \\
  \varsigma(1), & \mbox{when coupling with $1$ in $\triangle \partial_\mu$.}
  \end{array} \right.
\end{equation}
Thus the biframe bundle of $M$ is
\begin{equation}\label{B.m}
  BF(M)=\{(\triangle \varsigma (x), \triangle e_\mu)\}=\{(\varsigma(x)\otimes \varsigma(1)\oplus \varsigma(1)\otimes \varsigma(x), e_\mu\otimes e \oplus e\otimes e_\mu )\},
\end{equation}
where $\mu \in \{0,1,2,3\}$, $e_\mu = \partial_\mu$, and in usual conditions $e=1$. We let $\Pi: BP(M)\rightarrow x$ be the nature projection, and let $\kappa_x$ map the local fiber $\Pi^{-1}(x)=G(x)\times G$ of the principal bundle $BP(M)$ to the standard fiber $G$. One can directly check that $\kappa_x$ ($x\in U$, where $U$ is a neighborhood of $M$) is the projection: $\triangle \varsigma \mapsto \varsigma(1)$. Let $U_\alpha$ and $U_\beta$ be two open neighborhoods of $M$ such that $U_\alpha \bigcap U_\beta \neq \varnothing$, and let $x\in U_\alpha \bigcap U_\beta$. For any $u \in \Pi^{-1}(x)$, let $\kappa_{\beta, x}u= \varsigma(1)$, and $\kappa_{\alpha, x}u= \varsigma'(1)$. Since $\kappa_x$ is the nature projection, we have $\kappa_{\beta, x} = \kappa_{\alpha, x}$. Thus $\lambda_{\alpha \beta}= \kappa_{\alpha, x}\kappa_{\beta, x}^{-1} \subset G$ is the identity transformation function, and the consistency of the bundle $BP(M)$ is spontaneously fulfilled.

Now we extend other ingredients of the spectral triple to
\begin{eqnarray}
&& \triangle \xi=\xi\otimes \xi, \label{bi.xi} \\
&& \triangle \gamma_M=\frac{1}{2}(\gamma_M\otimes 1 + 1\otimes \gamma_M), \label{bi.gamma} \\
&& \triangle J_M = \sqrt{-1} J_M\otimes J_M. \label{bi.j}
\end{eqnarray}
$\xi$ in (\ref{bi.xi}) is defined similarly as $\varsigma$:
\begin{equation}\label{xi.x.1}
  \xi=\left\{\begin{array}{cc}
  \xi(x) \in L^2(M,S), & \mbox{when coupling with $\partial_\mu$ in $\triangle \partial_\mu$,} \\
  \xi(1)\equiv \varphi_x'\xi(x), & \mbox{when coupling with $1$ in $\triangle \partial_\mu$.}
  \end{array} \right.
\end{equation}
In this work, we restrict $C^\infty(M)$ and $L^2(M,S)$ to the subalgebra and subspace, such that $(\gamma_M\otimes 1)(\triangle \varsigma)=(1\otimes \gamma_M)(\triangle \varsigma)$ and $(\gamma_M\otimes 1)(\triangle \xi)=(1\otimes \gamma_M)(\triangle \xi)$ hold. Then $(\triangle \gamma_M)$ is still a $\mathbb{Z}/2$-grading.
The tensor products in the extension, e.g., $\varsigma \otimes \varsigma$, are divided into two parts restricted to the two frames, e.g., $\varsigma(x) \otimes \varsigma(1)\oplus \varsigma(1) \otimes \varsigma(x)$. Here $\varsigma(x)$ is the part coupling with $\partial\!\!\!/_M$ of $\triangle \partial\!\!\!/_M$ and thus is the part defined in the coordinate frame, and $\varsigma(1)$ is the part coupling with ``1'' of $\triangle \partial\!\!\!/_M$ and thus is the part defined in the non-coordinate frame. Then we let the scalar multiplications on fields in the two frames be independent to each other, i.e.,
\begin{eqnarray}
  k_c(\varsigma \otimes \varsigma) &=& (k_c\varsigma(x))\otimes \varsigma(1) \oplus \varsigma(1)\otimes (k_c\varsigma(x)), \label{kc} \\
  k_n(\varsigma \otimes \varsigma) &=& \varsigma(x)\otimes (k_n \varsigma(1)) \oplus (k_n \varsigma(1))\otimes \varsigma(x). \label{kn}
\end{eqnarray}
Similar equations hold for $\xi\otimes \xi$. In this sense, the part of the spectral triple restricted to the coordinate frame or non-coordinate frame has a linear structure.

One can directly check that the $K$-theoretic dimension (i.e., 10 modulo 8) of $M\times F$ is the same with that in the unextended case, and thus the fermion doubling problem~\cite{Lizzi:1996vr,nc.rev,con} is avoided.

In the original unextended geometry~\cite{con,nc.rev,Connes:1988ym,Chamseddine:1996zu,Chamseddine:1991qh}, gravity and the bosonic couplings are characterized by the spectral action Tr$(f(D/\Lambda))$, where $f$ is any positive even function and $\Lambda$ is the parameter fixing the mass scale. This spectral action can be calculated by the coefficients $a_{2n}$ ($n\in\{0,1,2\}$) of the heat expansion Tr$(e^{-tD^2})\sim \sum a_\alpha t^\alpha$ ($t\rightarrow 0$), while $a_{2n}$ are determined by traces of the gauge potentials and curvature tensors of the geometry~\cite{con,nc.rev,Connes:1988ym,Chamseddine:1996zu,Chamseddine:1991qh}. The spectral action in our biframe extension is still written as Tr$(f(D/\Lambda))$, where $D=\triangle \partial\!\!\!/_M\otimes 1 + \triangle \gamma_M\otimes D_F$. Expressions of $a_{2n}$ are formally the same with those in the unextended geometry. Now we show that the gauge potentials and the curvature tensors in $a_{2n}$ keep unchanged under the extension.

In the original  geometric paradigm, the inner fluctuation of the metric replaces $D$ by $D+A+JAJ^{-1}$, where $A=\sum_i a_i[D, a_i']$ ($a_i, a_i'\in \mathcal{A}$) is the gauge potential containing the representations of gauge bosons and Higgs~\cite{Chamseddine:1991qh,Chamseddine:2010ud}. In our biframe extension, $A$ is rewritten as
\begin{eqnarray}
  A&=&\sum_i \triangle\varsigma_i\otimes a_{Fi} [(\triangle\partial\!\!\!/_M\otimes 1+ \triangle\gamma_M \otimes D_F), \triangle\varsigma_i'\otimes a_{Fi}'] \nonumber \\
  &=& \sum_i\frac{1}{2}(\varsigma_i(x)\otimes a_{Fi} [\partial\!\!\!/_M\otimes 1, \varsigma_i'(x)\otimes a_{Fi}']\otimes (\varsigma_i(1)\varsigma_i'(1))\nonumber \\
  & & +  (\varsigma_i(1)\varsigma_i'(1))\otimes  \varsigma_i(x)\otimes a_{Fi} [\partial\!\!\!/_M\otimes 1, \varsigma_i'(x)\otimes a_{Fi}'] ) \nonumber \\
  &=&\frac{1}{2}(A_0\otimes \mathcal{L}_1+ \mathcal{L}_1\otimes A_0),
\end{eqnarray}
where $\varsigma, \varsigma' \in C^\infty(M,S)$, $a_F, a_F'\in \mathcal{A}_F$, $\mathcal{L}_1=\sum_i \varsigma_i(1)\varsigma_i'(1)=\sum_i \varsigma_i''(1)$ ($\varsigma''=\varsigma\varsigma'$) is the factor in the non-coordinate frame, and $A_0$ is the gauge potential in the coordinate frame and is formally the same with $A$ in the unextended geometry. $\mathcal{L}_1$ is the extra factor coming from our extension. Since this factor does not specify extra gauge transformations, $A$ is parameterized in the same way of that for $A_0$. In other words, the representations of bosons in the inner fluctuation of the geometry are not changed by our biframe extension.

Now we consider the metric tensor $g$ that maps each pair of vectors of a tangent bundle to a real number. We let $\xi^\mu(x)$ be valued in tangent vectors of $M$, and give the corresponding square of the line element by
\begin{eqnarray}
ds^2&=&\sum_{\mu, \nu}g_{\mu\nu} (\xi^\mu(x)\otimes \xi(1)\oplus \xi(1)\otimes \xi^\mu(x))\nonumber \\
& &\otimes(\xi^\nu(x)\otimes \xi(1)\oplus \xi(1)\otimes \xi^\nu(x))
\nonumber \\
&=& \sum_{\mu,\nu}\mathfrak{g}_{\mu\nu}\xi^\mu(x)\otimes \xi^\nu(x),
\end{eqnarray}
where $\mathfrak{g}_{\mu\nu}=g_{\mu \nu} \otimes \mathcal{L}_2\oplus\mathcal{L}_2\otimes g_{\mu \nu}$ and $\mathcal{L}_2=\sum \xi(1)\xi(1)$. In the case that $\xi^\mu(x)$ is valued in $\{dx^\mu\}$, $g_{\mu\nu}$ is exactly the the metric tensor in general relativity. Then all curvature tensors in the spectral triple are multiplied by global factors with dependence on the tensor products of $\mathcal{L}_2$. The traces of the extra factors are real scalar coefficients in the coordinate frame corresponding to the ordinary frame bundle $F(M)$ and can be removed by a reparameterization. The spectral action consequently keeps unchanged under the biframe extension. The contributions of extra factors $\xi(1)$ in $\triangle\xi$ and $\mathcal{L}_1$ in $A$ to the fermionic part of the action are scalar coefficients, which can be absorbed by coupling coefficients without changing the action. Thus, the action of the geometry keeps invariant under the extension.

\section{Quaternion extension on the geometry $M$}
\label{sec2}

Now we extend the expressions of fields by complexified quaternion numbers. We represent the real unit $I_0$ and imaginary units $I_i$ ($i\in \{1,2,3\}$) of the extended quaternion space by $4\times 4$ matrices. Then we have $I_\mu^2=\eta_{\mu\mu}$I$_4$, where I$_4$ is the 4$\times$4 identity matrix and $\eta=||\eta_{\mu\nu}||=$diag$(+1, -1, -1, -1)$.

We distinguish the left and right quaternion multiplications. The left multiplication rules of $I_k$ are equal to those of $\sigma^{ij}$ with $i,j,k\in \{1,2,3\}$ and $\epsilon_{ijk}=+1$. Here $\sigma$ is determined by gamma matrices: $\sigma^{\mu\nu}=\frac{1}{2}[\gamma^\mu,\gamma^\nu]$ ($\mu,\nu\in \{0,1,2,3\}$). The right multiplications of imaginary units of the quaternion space are similarly represented by products of $\{\sigma^{0i}, \sigma^{j0}\}$. Thus, the left and right multiplications of the quaternion space can be represented by the basis $\{\sigma^{ij},\sigma^{0i},\sigma^{j0}\}$. Since $\{\sigma^{\mu\nu}\}$ is a representation of the Lorentz group, the extended fields are equivalent to the usual complex fields governed by Lorentz transformations~\cite{Fischer:1940qza,Synge:1972zz,Teli:1980mr,Dixon,DeLeo:1998ye,Morita:2007vc,DeLeo:2000ik}. Since such fields in non-coordinate frame can be valued in $\gamma^a$ ($a\in\{0,1,2,3\}$), we can define $\gamma^a$ as the basis of the vector space $E$. Thus, combined with the quaternion extension, the extra non-coordinate frame of the biframe bundle is no longer a hidden frame.

Since the basis vectors of $E(x)$ are $\{dx^\mu\}$ and those of $E$ are $\{\gamma^a\}$, we need to distinguish the Lorentz transformations in the two frames. The Lorentz transformation in the coordinate frame is the ordinary Lorentz transformation $L_\mu^{\,\,\,\nu}$ in the coordinate spacetime. Without the biframe extension, the quaternion extended fields in the frame bundle $F(M)$ are equal to those in the ordinary noncommutative geometry. The Lorentz transformation in the non-coordinate frame is denoted as $\Lambda_a^{\,\,\,b}\in\mathrm{SO}(1,3)\cong\mathrm{SP}(1,3)$ (and replace the basis $\{\sigma^{\mu\nu}\}$ by $\{\sigma^{ab}\}$ with $a,b\in\{0,1,2,3\}$). Thus, the quaternion extended non-coordinate frame is equivalently governed by spinnic Lorentz transformations. We let the coordinate frame be globally flat, and then this frame is equal to the coordinate spacetime frame in the biframe theory of Ref.~\cite{wu.qg}. We let the non-coordinate frame be flat at every point of the bundle, and then this frame is equal to the non-coordinate gravity spacetime frame in Ref.~\cite{wu.qg}. Hence, the two frames of the bundle under our extensions on the geometry establish the biframe spacetime which can describe gravity in the way of QFT.

The field $\triangle\xi^\mu(x)=\xi^\mu(x)\otimes \xi(1)\oplus \xi(1)\otimes \xi^\mu(x)$ is sided on both of the two frames, while $\xi^\mu(x)$ in $\triangle\xi(x)^\mu$ is the field restricted to the coordinate spacetime frame and $\xi(1)$ is restricted to the non-coordinate spacetime frame. $\xi(1)\equiv\xi_a$ is governed by the spinnic transformation $\Lambda_a^{\,\,\,b}$, which is a local gauge transformation at every point of the total space, and thus one can rewrite $\triangle\xi^\mu(x)$ as $\xi^{\,\,\,\mu}_a(x)$. The index $a$ can be lifted and lowered by $\Lambda_b^{\,\,\,a}$. The field $\xi^{\,\,\,\mu}_a$ can be valued in the basis $\{\frac{1}{2}\gamma_a\}$:
\begin{equation}
  \xi^\mu(x)= \frac{1}{2}\gamma^a \xi_a^{\,\,\,\mu}(x).
\end{equation}
In the case that $\xi^\mu(x)$ is valued in basis $\{dx^\mu\}$:
\begin{equation}
  \xi(x)=\xi^\mu(x) \partial_\mu,
\end{equation}
the field $\xi_a^{\,\,\,\mu}(x)$ is equivalent to the bicovariant vector field $\hat{\chi}_a^{\,\,\,\mu}$ which is the gravifield in the biframe spacetime in Ref.~\cite{wu.qg}.

Now we investigate the curvature tensors, and representations of bosons and fermions in the extended geometry. Since coordinate frame is globally flat and non-coordinate frame is flat at every point of the total space, the square of $ds$ reads
\begin{equation}
ds^2=\sum \eta_{ab} \xi^{a} \otimes \xi^{b},
\end{equation}
where $\xi^a=\Lambda^a_{\,\,\,b} \xi^b$. To make explicit comparison with general relativity, we investigate the metric in the coordinate basis $\{dx\}$. Notice that $\xi^a$ can be expanded into terms with basis $\{dx\}$ by $\xi^a= \xi^{a}_{\,\,\, \mu}(x) dx^\mu$. Then we have
\begin{eqnarray}
  ds^2 &=& \sum \eta_{ab} (\xi^{a}_{\,\,\, \mu}(x)dx^\mu) \otimes (\xi^{b}_{\,\,\, \nu}(x)dx^\nu) \nonumber \\
   &=& \sum \xi_{\mu\nu} dx^\mu\otimes dx^\nu,
\end{eqnarray}
which is in the similar form of the line element in Ref.~\cite{wu.qg}. Here $\xi_{\mu\nu}=\eta_{ab}\xi^a_{\,\,\,\mu}\otimes \xi^b_{\,\,\,\nu}$, which is in the similar form of the gravimetric field in Ref.~\cite{wu.qg}. $\xi_{\mu\nu}$ plays the role as the effective metric. Such effective metric is coordinate-dependent due to the fields $\xi^a_{\,\,\,\mu}$. Curvature tensors in the coordinate Minkowski spacetime can be consequently defined. The corresponding contributions of the curvature tensors to the spectral action are formally the same with those in the unextended geometry. Since the gravity effect can be encoded by curvatures characterized by the fields $\xi^a_{\,\,\,\mu}$ of extended $M$, we can call $\xi^a_{\,\,\,\mu}$ the gravifield.

As discussed above, the gauge potential from the inner fluctuation is $A=\frac{1}{2}(A_0\otimes \mathcal{L}_1+\mathcal{L}_1\otimes A_0)$. The original geometry determines the algebra $\mathcal{A}_F$ of the spectral triple of $F$ to be $\mathbb{C}\oplus\mathbb{H}\oplus M_3(\mathbb{C})$~\cite{Chamseddine:1991qh,nc.rev,Chamseddine:2010ud}. $A_0=\sum_i a_{Fi}[D_F, a_{Fi}']$ then generates the representation of the gauge symmetry group SU$(3)\times$SU$(2)\times$U$(1)$ of the standard model. In our extension, the field $\mathcal{L}_1=\sum_i\varsigma_i(1)$ restricted to the coordinate frame of the ordinary frame bundle is transformed under $\Lambda_a^{\,\,\,b}$, and can be valued in $\{\gamma^a\}$. Thus there is an extra index $\{a\}$ for $\varsigma_a$, $\mathcal{L}_{1a}$, and $(A_0)_a$. The gauge fields $\mathcal{A}_a$ in $(A_0)_a$ are correspondingly governed by the spinnic Lorentz transformation. Namely, there are non-coordinate gravity indexes $\{a,b\}$ for gauge fields. However, the gravity indexes of the curvature tensors and the gauge fields are contracted with indexes of $\eta_{ab}$ or $\eta^{ab}$ in the spectral action, then can be eliminated. Thus the contribution of the gravifield to Tr$(f(D_A/\Lambda))$ is the scaling effect $|\xi|^{2n}$ for the curvature tensors and gauge fields. We let the gravifield fulfill $|\xi|^2=1$. Then the spinnic Lorentz symmetry SP$(1,3)$ of the non-coordinate frame spacetime is a hidden symmetry in the gravity and bosonic parts of the spectral action.

The fermionic part of the action in the extended geometry is
\begin{eqnarray}
&&   \frac{1}{2}\langle (\triangle J_M\otimes J_F)(\triangle \xi \otimes \psi)|\triangle \partial\!\!\!/_M \otimes 1 + \triangle \gamma_5\otimes D_F|\triangle \xi \otimes \psi\rangle \nonumber \\
&&  = \frac{1}{2}\langle \bar{\xi}_a\gamma^a\otimes \bar{\psi}|\xi_a^{\,\,\,\mu}\gamma^a(\partial_\mu + i \otimes (D_A)_\mu) |\xi_a\gamma^a\otimes \psi\rangle,
\end{eqnarray}
where $\psi$ are the representations of fermions contained in $\mathcal{H}_F$ determined by the original geometry of $F$~\cite{con,nc.rev,Chamseddine:2006ep}. One can directly check that the fermionic part of the action is still explicitly coupled with gravifields $\xi_a^{\,\,\,\mu}$, even if we let $|\xi|^2=1$. Thus the spinnic symmetry SP$(1,3)$ of the gravity spacetime manifests itself as the global Lorentz symmetry in the fermionic part of the physical action. As discussed in previous works (see, e.g., Refs.~\cite{wu.qg,Chamseddine:2016pkx,Utiyama:1956sy,Kibble:1961ba,Sciama:1964wt,guo.wu.zhang,Yang:1974kj,Hehl.ge.spi,Ivanenko:1984vf,Hehl:1994ue}),  gravity fields governed by the SP$(1,3)$ symmetry can be quantized in a similar way of QFT. The quantization of $\xi_a^\mu$ decomposes the field into
\begin{equation} \label{sym.bre}
  \xi_a^{\,\,\,\mu}=\langle{\xi}_a^{\,\,\,\mu}\rangle + h_a^{\,\,\,\mu},
\end{equation}
where $\langle{\xi}_a^{\,\,\,\mu}\rangle$ is the background structure, and $h_a^{\,\,\,\mu}$ is the quantum gravifield. The quantization can be applied to the curvature tensors via the quantization of fields $\xi^{\,\,\,\mu}_{a}\otimes \xi^{\,\,\,\nu}_{b}$. Thus a unified quantum theory of gravity and the standard model is established in the geometric paradigm.

We mention that the unified theory in the geometric paradigm can also be constructed without the explicit appearance of curvature tensors in the spectral action. That is, we use the flat metric $\eta$ instead of the effective metric $g$, and let the spectral action be free of curvature tensors. Then we let any constant matrix noncommutative with fields $\xi(1)$ in some sense play the role as the Dirac operator in the non-coordiante spacetime frame. We let such matrix be $\sigma^{ab}$, and make biframe extension on $\sigma^{ab}$ by $\triangle \sigma^{ab}=\frac{1}{2}(\sigma^{ab}\otimes 1 + 1\otimes \sigma^{ab})$. The corresponding inner fluctuation can be written as $A_\sigma=\triangle\varsigma\otimes a_F [\triangle\sigma^{ab}\otimes 1, \triangle\varsigma'\otimes a_F']$. We mention that the part of field coupling with the nontrivial factor of $\triangle \sigma^{ab}$ is the one restricted to the non-coordinate frame. One can directly check that $A_\sigma$ is parameterized by fields governed by the spinnic SP$(1,3)$ gauge transformation. We denote these fields by $\Omega_{ab}^\mu$. Then we have
\begin{equation}
  \Omega^\mu = g_s \Omega^\mu_{ab}\frac{i}{4}\sigma^{ab},
\end{equation}
where $g_s$ is the coupling constant of the gauge field. We mention that the Dirac operator $\sigma^{ab}$ in the locally flat non-coordinate spacetime frame has explicit physical meaning. The rotation operation $\int_0^{2\pi} d\hat{\phi}$ (where $d\hat{\phi}$ is the operator that rotates a system by $d\phi$ angle) in the coordinate spacetime frame can be replaced by $\Pi_i\sigma^i=-$I$_4$ (where $\sigma^i\equiv\sigma^{jk}$ with $i,j,k\in\{1,2,3\}$ and $\epsilon_{ijk}=+1$) in the non-coordinate spacetime frame. An eigenstate under one such rotation gets a coefficient $-1$, and under two such rotations keeps invariant. Thus one can define the eigenstate of this rotation operator as a spinor. This argument explicitly shows that the non-coordinate spacetime frame is the spinnic spacetime frame.

The gauge tensor fields $\mathcal{F}_{ab}$ of $\mathcal{A}_a$ can be defined by $\sigma^{ab}$, which in the non-coordinate spacetime frame is the analogue of the Dirac operator in the coordinate spacetime frame. Reparameterize $\sigma^{ab}$ by the antisymmetric part of tensor square of $\gamma^a$:  $\gamma^a\otimes \gamma^b /(\gamma^c\otimes \gamma^c) \sim \xi^a\wedge\xi^b$. Then we have $\mathcal{F}= \mathcal{F}_{ab}\xi^a\wedge\xi^b$.

Now the spectral action Tr$(f(D_A/\Lambda))$ only contains gauge couplings of $\Omega_{ab}^{\mu}$ and $\mathcal{F}_{ab}^{\mu\nu}$. The gravity effect is characterized by the gauge couplings of $\mathcal{F}_{ab}$ and $\Omega_{ab}$. The technical details of such gravity couplings are described by the gravity action in Ref.~\cite{wu.qg}, but without the local scaling gauge symmetry. One can let the extended geometry be locally scaling gauge invariant, then corresponding gravity fields in Ref.~\cite{wu.qg} can be generated.

We mention that the gauge formulation of gravity is originally initiated by Ref.~\cite{Utiyama:1956sy}, and this gauge formulation is one start point of our extended geometry and the biframe theory~\cite{wu.qg}. The Einstein-Cartan gravity theory, as another application of the gauge formulation of gravity, uses the general asymmetric affine connection instead of the Levi-Civita connection to characterize the coupling of spin matters and gravity. Since the spin in gravity is concerned, both our extended geometry (and the corresponding biframe gravity) and the Einstein-Cartan theory~\cite{Cartan:1922,Cartan:1923,Cartan:1924,Cartan:1925,Kibble:1961ba,Sciama:1964wt,Hehl.ge.spi} give torsional gravity. These two theories have more similarities. For instance, the non-coordinate spacetime frame in the biframe gravity is a generalization of the vielbeins in original Einstein-Cartan theory. The vielbeins, which formulate the affine connection, come from distinguishing the tangent space of the original spacetime from the tangent space of an associated affine space. This theory can recover general relativity by torsion-free condition, usually as $\nabla g=0$ (where $\nabla$ is the covariant derivative corresponding to the affine connection and $g$ is the metric tensor field). In the biframe gravity, the space corresponding to the non-coordinate frame with the basis $\{\frac{1}{2}\gamma_a\}$ is equipped by less structures than that of the vielbeins. The parallel transport and corresponding affine connection are not explicitly given in the space, and the covariant derivative of tangent fields valued in $\{\frac{1}{2}\gamma_a\}$ is not explicitly defined either. Thus, as mentioned in Ref.~\cite{wu.qg}, torsional effect in the biframe gravity can not be directly removed. In other words, the biframe gravity constructs vielbeins with less conditions on the corresponding space and gives, somehow, more ``radical'' torsional effect of gravity. Thus, whether the biframe gravity can shed a light on the understanding and the application of the torsional effect of gravity is an open issue.

As discussed in Ref.~\cite{wu.qg}, the biframe gravity can recover the Einstein general relativity after symmetry breaking (\ref{sym.bre}) in the low energy limit. The general relativity is a low energy effective theory of the biframe gravity, while the symmetry breaks into the hidden symmetry GL$(4,R)$. In our extended geometry, how the gravity recover the general relativity can be understand from other aspects. As discussed above, one can formulate the analogue of the Dirac operator in the non-coordinate spacetime frame to give the spectral action. Similar analogue of the covariant derivative in the space corresponding to the non-coordinate frame can also be constructed, and thus one can reformulate the corresponding ``affine connection'' in the form of the Einstein-Cartan theory. The corresponding torsion-free condition can also be given in this form without introducing extra derivative terms (since the analogues are not exactly derivative operators). Moreover, one can directly transmute the space of the non-coordinate frame into an affine space. For instance, for every field $\xi_a$ in the space $X$ corresponding to the non-coordinate frame, one can consider different values of $\langle{\xi}_a\rangle$ in (\ref{sym.bre}) as variants (before symmetry breaking) and include them into space $X$ and then consider the field $h_a=\xi_a- \langle{\xi}_a\rangle$ as a vector (or translation) in a space $V$. Then for any variant $\langle{\xi}_a\rangle \in X$ and any $h_a \in V$ there is a unique point $\xi_a \in X$ such that $\xi_a-\langle{\xi}_a\rangle= h_a$. Thus an affine space is defined. One can accordingly give the affine connection and construct the extended geometry in the formulation of Einstein-Cartan theory. In this sense, our extended geometry can be a generalization of the Einstein-Cartan theory.

\section{Conclusion}
\label{sec3}

We refer the field $\xi$ defined in the extended space $M$ as the gravifield, and characterize the gravity effect in the biframe spacetime. One can adopt effective curvature tensor as the flat metric tensor coupled with the gravifield. One can also characterize the gravity effect by gauge fields coming from inner fluctuations of the ``Dirac operator'' in the non-coordinate spacetime frame. The spectral action in the latter viewpoint is similar with that in Ref.~\cite{wu.qg}, where the physical notions specifying gravity are not curvatures but directly the gauge fields.
In the extended geometry, the spinnic gauge symmetry of gravifields is hidden in the gravity and bosonic part of the action, and explicitly appears in the fermionic part of the action. This is the same with the conclusion of~Ref.\cite{wu.qg} after symmetry breaking.

In summary, we provide a geometric way to formulate the biframe gravity theory of Ref.~\cite{wu.qg}, and then geometrically unify gravity and quantum field theory (QFT) from the extended noncommutative geometry. We construct a globally flat coordinate spacetime frame by the coordinate part of the biframe bundle for the QFT. Fields of the standard model are defined in such spacetime frame and can be quantized as in the QFT. We then construct a locally flat non-coordinate spacetime frame by the non-coordinate part of the biframe bundle expressed by quaternion numbers for the gravity effect. The field defined in both frames of the biframe spacetime is the gravity field, and can be quantized in such ``flat'' biframe spacetime. This gravity field couples with all the fields in the coordinate frame to make the fields of the standard model meet general relativity. Thus, from a mathematical viewpoint, we arrive at a geometric framework in which both gravity and quantum fields of the standard model are quantized and also quantum fields of the standard model are coupled with the gravity. Gravity and the QFT then can be studied in an unified manner in such geometric paradigm.

\section*{Acknowledgments}

This work is supported by the National Natural Science Foundation of China (Grants No.~11475006 and No.~11120101004).

\bibliographystyle{iopart-num}

\end{document}